\documentclass[a4paper]{article}

\usepackage[pages=all, color=black, position={current page.south}, placement=bottom, scale=1, opacity=1, vshift=5mm]{background}
\SetBgContents{
	\tt This work is shared under a \href{https://creativecommons.org/licenses/by-nc-nd/4.0/}{CC BY-NC-ND 4.0 license} unless otherwise noted
}      

\usepackage[margin=1in]{geometry} 

\usepackage{amsmath}
\usepackage{amsthm}
\usepackage{amssymb}

\usepackage[utf8]{inputenc}
\usepackage{hyperref}
\hypersetup{
	unicode,
	pdfauthor={Author One, Author Two, Author Three},
	pdftitle={A simple article template},
	pdfsubject={A simple article template},
	pdfkeywords={article, template, simple},
	pdfproducer={LaTeX},
	pdfcreator={pdflatex}
}

\usepackage[sort&compress,numbers,square]{natbib}
\theoremstyle{plain}

\theoremstyle{definition}

\usepackage{graphicx, color}
\usepackage{subcaption}
\usepackage{booktabs}
\graphicspath{{fig/}}

\usepackage{algorithm, algpseudocode} 
\usepackage{mathrsfs} 

\usepackage{lipsum}

\title{Balancing Turnover and Promotion Outcomes: Evidence on the Optimal Hybrid-Work Frequency}
\author{Xuan Lu$^1$\and Yulin Yu$^{2,3}$}

\date{
	$^1$College of Information Science, University of Arizona \\ \texttt{luxuan@arizona.edu}\\%
	$^2$School of Information, University of Michigan \\ 
    $^3$Northwestern Institute on Complex Systems, Northwestern University \\\texttt{yulinyu@umich.edu}\\[2ex]%
}

\begin{document}
\maketitle

\begin{abstract}
Hybrid work policy, especially return-to-office requirements, remains a globally salient topic as workers, companies, and governments continue to debate and disagree. Despite extensive discussions on the benefits and drawbacks of remote and hybrid arrangements, the optimal number of remote days that jointly considers multiple organizational outcomes has not been empirically established. Focusing on two critical career outcomes—turnover risk and promotion—we examine how remote work frequency shapes employee trajectories using large-scale observational activity data from a company with over one million employees. We find that increased remote-work frequency is associated with an initial decrease and then an increase in turnover, while promotion likelihood initially rises and then declines. Accordingly, we identify approximately two remote days per week as an optimal balance—maximizing promotion, a positive outcome for employees, while minimizing turnover, which is undesirable for organizations and may indicate negative employee experiences.
These patterns vary across subgroups defined by gender, role type, and leadership status. Several notable results emerge. First, male employees derive greater promotion benefits from remote work than female employees. Second, support workers (non–core business roles) do not experience promotion gains, and the reduction in turnover at their optimal remote-work frequency is marginal compared with employees in core business roles. Third, organizational leaders face greater challenges in remote settings than individual contributors: their turnover risk increases substantially at higher remote frequencies, and their likelihood of promotion decreases as remote frequency rises. 
We further show that time-allocation patterns partly explain how remote-work frequency influences these career outcomes.\\
    
\end{abstract}


\section{Introduction}
Remote work remains a persistent global topic~\cite{Aksoy2025-pr, barrero2021working, malone2004future}, with workers, companies, and governments continuing to hold significantly different views on whether—and how—it should be adopted. From an employee perspective, remote work~\cite{bloom2015does, Bloom2024-yi, barrero2021working, dua2022americans, Schall2019-gn} offers benefits like flexibility and autonomy, and thus workers tend to favor it~\cite{deci2004handbook, jung2011understanding}. Companies, on the other hand, recognize both the advantages and drawbacks: in-person work can strengthen organizational culture and team cohesion, while remote work can reduce facility costs. Industry policies remain mixed: some companies~\cite{megan24companies} continue to offer fully remote options, whereas others have tightened their policies and implemented return-to-office mandates~\cite{andy24message}. Governments, by contrast, often adopt a more cautious or negative stance toward promoting remote work due to broader economic~\cite{Ramani2024-ju} and social concerns, such as downtown economic recovery, reduced demand for commercial real estate, and public service performance. For example, in January 2025, the U.S. government announced that federal agencies would phase out remote work arrangements~\cite{white25order}. City governments such as San Francisco~\cite{sf24standard} have also publicly encouraged employers to bring workers back to offices to counter rising commercial real estate vacancies and declining urban business activity.

These differing opinions among stakeholders may stem from the perceived double-edged effects of remote work, leading to the adoption of hybrid work arrangements. On the one hand, remote work can improve productivity~\cite{Bloom2024-yi}, enhance work-life balance~\cite{Babapour_Chafi2021-so}, and increase perceived autonomy~\cite{Gajendran2024-be} and job satisfaction~\cite{Schall2019-gn}. On the other hand, remote work may reduce productivity in certain settings~\cite{Monteiro2019-rn}, make collaboration networks more static and siloed~\cite{Yang2021-nn, Yang2022-ki,yu2023large}, decelerate the generation of creative ideas~\cite{Brucks2022-tx} and breakthrough discoveries~\cite{lin2023remote}, and worsen emotional well-being~\cite{du-Plooy-and-Gert-Roodt2010-zs}. Moreover, frequent remote work may blur the boundaries between work and life~\cite{Hayashi2024-lp}. As a result, many organizations now combine remote and in-person work through hybrid models~\cite{prarthana23google, david23meta, necole25new, catherine22snap}. However, the implementation of hybrid work policies varies across organizations. For example, in the IT industry, Google~\cite{prarthana23google} and Meta~\cite{david23meta} require employees to be in the office at least three days per week, whereas Intel~\cite{necole25new} and Snap~\cite{catherine22snap} mandate a minimum of four days. 

The wide variation in work-from-home policies might reflect the lack of clear evidence on what arrangement yields the greatest benefits or optimal outcomes. While numerous studies have confirmed the benefits of hybrid work~\cite{Bloom2024-yi, Babapour_Chafi2021-so, Gajendran2024-be, Schall2019-gn}, we still lack a clear understanding of its optimal implementation—even for the simplest question: \textit{What is the optimal hybrid work strategy, particularly regarding the frequency of remote work (hereafter referred to as \textit{remote frequency}), that balances work-related outcomes and maximizes its benefits?} 

The answer to this question can provide valuable evidence to help align or balance the interests of different stakeholders. For employees, an optimal frequency of remote work could help maintain effective connections with co-workers and achieve an appropriate balance between work and life. For employers, well-informed organizational policies could enhance employee motivation, productivity, and job satisfaction, thereby contributing to the long-term success of the company. Additionally, researchers in organizational behavior, labor economics, and future-of-work studies can build on this strong empirical foundation when designing surveys, interviews, data analyses, and experiments.

To determine the optimal frequency of remote work, we examine its relationship with two key workplace outcomes: turnover and promotion.\textit{Turnover} is an indicator of subjective career success~\cite{Judge1995-xa, Ng2005-xj,gamba2024exit} and has had a mixed relationship with remote work. \cite{Schall2019-gn} shows that remote work intensity is positively correlated with job satisfaction, a key factor influencing turnover intentions. 
\cite{Bloom2024-yi} demonstrates in a randomized experiment that two days of remote work per week can significantly improve job satisfaction and reduce turnover risk by one-third. Remote work, however, is also accompanied by challenges, such as ineffective communication and low engagement levels~\cite{schawbel2018survey, silva2024perceived, santana2023influence}, which can contribute to higher turnover risks. Moreover, the increased mobility associated with remote work may motivate workers to pursue better job opportunities. 
The other outcome that we study, \textit{promotion}, is an indicator of objective career success~\cite{Judge1995-xa} and has also shown varying relationships with remote work.  
For example, \cite{Bloom2024-yi} finds no significant difference in promotion between onsite workers and those who work remotely two days per week; \cite{Golden2020-ju} reports a negative correlation between the extent of remote work and promotions; and \cite{Golden2019-ff} shows a positive association between the extent of remote work and job performance. These two outcomes are important not only for employees but also for the long-term development of companies and, more broadly, to society. Given the complex dynamics between remote work and career success outcomes, this study aims to explore whether there exists an optimal frequency of remote work that benefits both outcomes. In addition, we examine the underlying mechanisms of the effects of remote work frequency by considering employees’ backgrounds and time-allocation behaviors.

A key challenge for this kind of study lies in the limited availability of large-scale, real-world datasets that capture diverse remote work patterns and their associated organizational outcomes. Using digitally traced workplace activity data from a Chinese technology education company, we obtained attendance records for 41,579 employees, reflecting remote work frequencies ranging from zero to five days per week, as well as demographic and organization-related characteristics, including promotion and turnover. Our study shows a U-shaped relationship between remote frequency and  turnover, and an inverted U-shaped relationship between remote frequency and promotion outcomes. 
Although the relationships between remote frequency and promotion and turnover vary, both outcomes indicate that hybrid work offers advantages over fully on-site or fully remote setups, with a beneficial range of up to three remote days per week. Within this range, hybrid work can leverage the benefits of remote work while mitigating the challenges associated with intensive remote arrangements. Furthermore, our results indicate that two days of remote work per week (more precisely, between 1.6 and 2.1 days) best balance the two outcomes. The effects of remote frequency can be moderated by employees’ gender, position type, and leadership status. The way workers allocate their time, measured in terms of working hours, working time flexibility, and after-hours work, mediates the effects of remote work frequency. This work advances research on how remote work frequency influences work-related outcomes and provides practical implications for employers, employees, and policymakers.

\section{Effects of Remote Frequency on Turnover}

We analyze remote work frequency in April 2021 and its effect in May 2021 to minimize the influence of external factors, such as the COVID-19 pandemic and the Double Reduction Policy, a Chinese education policy issued on July 24, 2021. As this policy may have had a significant impact on the company's organizational structures, we chose to exclude the period affected by the policy. See~\nameref{sec:methods} for details. The year will be omitted hereafter for brevity. 
To investigate whether the frequency of remote work in April influences turnover risk in May, we select employees with continuous employment who were active at least one month before the study period, i.e., those who were active both March and April, as new hires tend to have higher turnover rates due to difficulties adjusting to the job~\cite{hom2017one}. This results in a population of 41,579 employees with an overall turnover rate of 12.3\% in May. 
Then, we perform a weighted OLS regression of turnover on remote frequency, where the weights are used to adjust for confounders. This approach balances covariates across different levels of remote work frequency, thereby approximating a randomized setting. The method and the list of confounders are described in~\nameref{sec:methods}. Furthermore, as a non-linear relationship between remote frequency and turnover is observed from data, we add a quadratic term of remote frequency in the regression. With the regression fitted with all significant coefficients ($p < 0.001$), we then visualize the relationship between remote frequency and turnover risk with a dose-response curve. The regression results are presented in Table S1 in Appendix.

\subsection{Turnover declines with increases in remote frequency up to 1.6 days per week, then rises}
\begin{figure*}[t!]
\centering
    \begin{subfigure}[b]{0.3\textwidth}
         \centering
         \includegraphics[width=\textwidth]{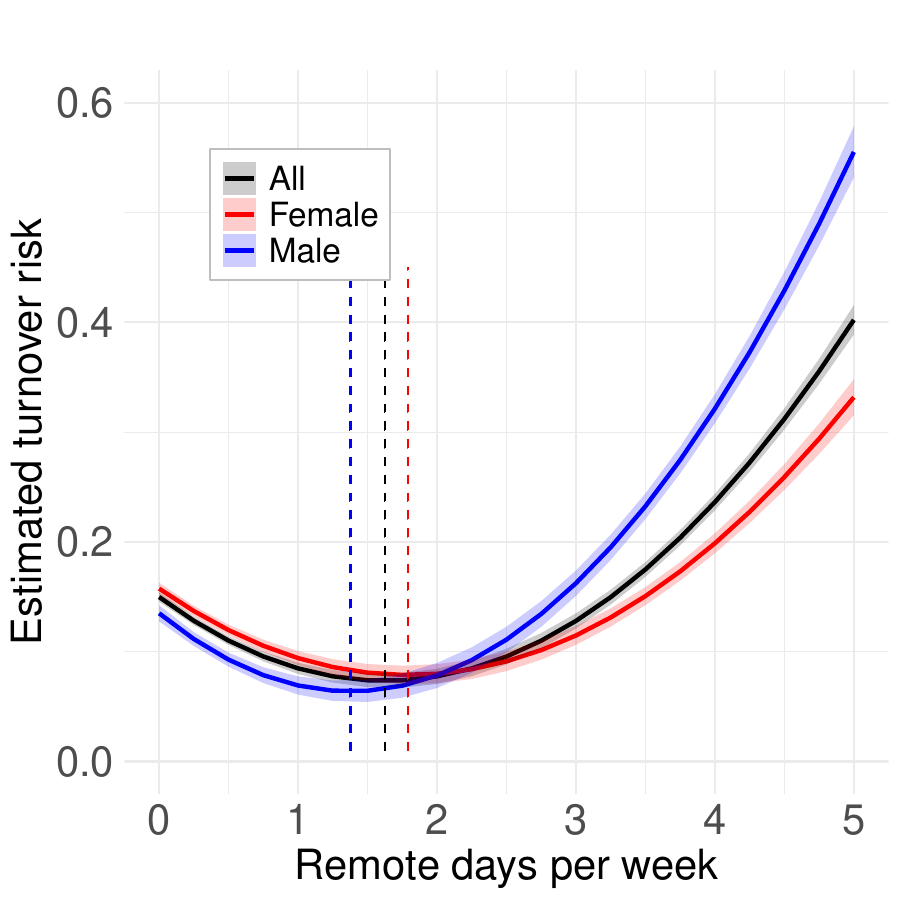}
         \caption{Gender}
         \label{fig:april_dropout_gender}
     \end{subfigure}
    \begin{subfigure}[b]{0.3\textwidth}
         \centering
         \includegraphics[width=\textwidth]{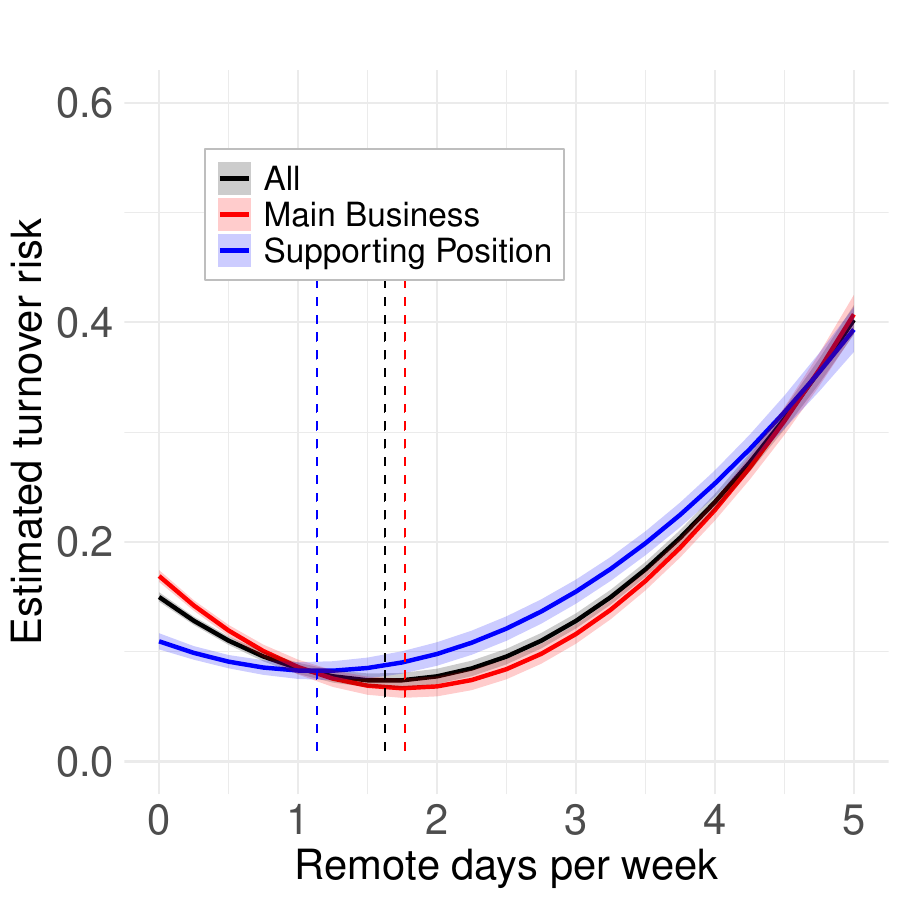}
         \caption{Position type}
         \label{fig:april_dropout_position}
     \end{subfigure}
     \begin{subfigure}[b]{0.3\textwidth}
         \centering
         \includegraphics[width=\textwidth]{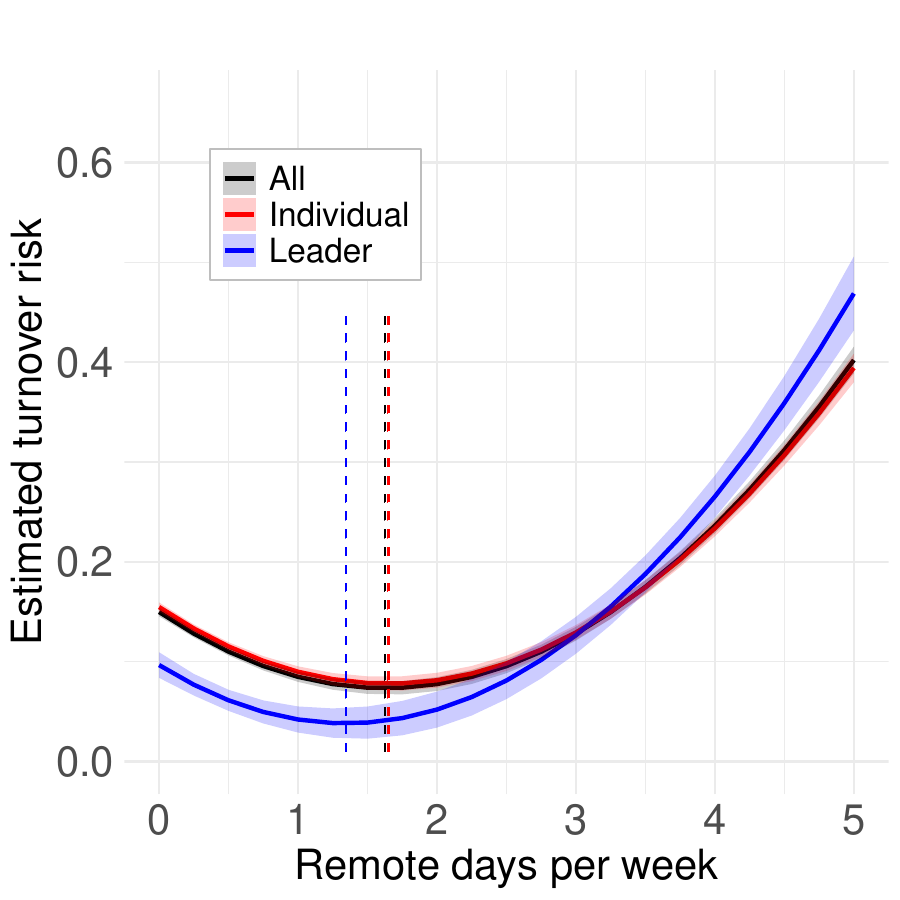}
         \caption{Leadership}
         \label{fig:april_dropout_leadership}
     \end{subfigure}
\caption{\textbf{Dose-response curves with 95\% confidence intervals for the weighted regressions of turnover risk on remote frequency.} As remote frequency increases, turnover risk initially decreases, reaching its lowest point at 1.6 remote days per week, and then begins to rise. Regressions on subgroups defined by gender, position type, and leadership status reveal heterogeneous effects. \textbf{(a)}: The optimal remote frequency for minimizing turnover risk is 1.4 days per week for men and 1.8 days per week for women. \textbf{(b)}: For main business workers, the optimal frequency is 1.8 days per week, while for support workers it is 1.1 days per week. See the definitions of position types in the main body.  \textbf{(c)}: For individual contributors, the optimal frequency is 1.7 days per week, whereas for leaders it is 1.3 days per week.}\label{fig:april_dropout}
\end{figure*}

Fig.~\ref{fig:april_dropout} illustrates the dose-response curve between remote frequency and turnover risk. When employees work fully onsite, the turnover risk is 15.0\%, indicating a higher-than-average likelihood of leaving the company (the overall turnover rate is 12.3\%) with other factors controlled. As remote frequency increases from zero to 1.6 days per week, the U-shaped curve indicates a reduction in turnover risk from 15.0\% to 7.4\%, a decrease of about one half, highlighting the benefits of remote work. However, beyond 1.6 days per week, turnover risk rises sharply and at 3.3 days per week surpasses that of fully onsite workers. It ultimately reaches 40.2\% for fully remote employees, nearly three times the level observed for fully onsite workers, suggesting that the challenges of remote work may outweigh its benefits. This result suggests both an optimal frequency of 1.6 days per week for minimizing turnover risk, as well as a beneficial range of remote frequency (greater than zero but less than 3.3 days per week), within which employees are less likely to leave the company. 
This range varies slightly across groups, as will be discussed shortly. Higher remote frequencies are associated with significantly increased turnover risk and should therefore be avoided or, at the very least, carefully examined to understand the underlying causes.

\subsection{Effects of remote work frequency on turnover are moderated by gender, position, and leadership}
The quadratic relationship between remote frequency and turnover risk holds across groups defined by gender, position type, and leadership role, but shows slight variations. Positions in our dataset are categorized into two types. The main business positions include all teaching and course-related research roles, while support positions cover marketing, operations, sales, and technical support roles. For leadership roles, employees with at least one direct report are classified as leaders, while the others are considered individual contributors. 

\subsubsection{Male turnover bottoms out at a lower remote frequency} 
Gender is an important factor in workplace performance. For example, women are less likely to enter competitions~\cite{He2021-ki}, less likely to request deadline extensions when they are adjustable~\cite{Whillans2021-uq}, and are often socially excluded in fields dominated by men~\cite{Cyr2021-ur}. In the context of remote work, gender can serve as an important moderator of work-related outcomes~\cite{barrero2021working, Bloom2024-yi}. Although existing studies have not directly examined how remote work frequency affects turnover risks across genders, Bloom et al.~\cite{Bloom2024-yi} show that, compared to working in the office, females experience a significant decrease in turnover when working remotely two days a week, while the decrease for males is insignificant—indicating that there are more favorable remote work outcomes for female workers. Our analysis, however, suggests that the main gender differences in the effects of remote work emerge at higher remote frequencies, where employees may face more challenges from remote work than benefits. More specifically, as shown in Fig.~\ref{fig:april_dropout_gender}, the difference in turnover risk between males and females remains stable at low levels of remote frequency (i.e., 0 to 1 day per week). However, males begin to exhibit an increasing turnover risk beyond a frequency of 1.4 days, surpassing that of females at around 2 days per week, followed by a sharp increase, reaching a turnover risk of 55.5\% when fully remote. Although females show a similar pattern of increasing turnover risk at higher remote frequencies, their turning point occurs at a frequency of 1.8 days—about half a day more than males. When females work fully remotely, their turnover risk is 33.2\%, which is 40.2\% lower than that of males. The difference between genders in turning points, as well as the differing rates of increase after those points, suggests that females and males adapt to remote work differently and encounter distinct challenges.

\subsubsection{Main business workers benefit more from remote work than support workers in reducing turnover} 
Fully onsite, main business workers have a 54.3\% higher turnover risk compared to support workers. However, as remote frequency increases, this risk reverses. Main business workers experience a significant decline in turnover risk, reaching its lowest of 6.7\% at 1.8 days per week.  
In contrast, support workers see a weaker benefit from remote work and face an increased turnover risk around a remote frequency of 1.1.  
This pattern indicates that main business workers benefit from a wider range of remote work frequencies associated with reduced turnover risk, and they experience a stronger effect at the optimal remote frequency, with a deduction of 60.5\% from onsite at lowest turnover risk compared to 24.7\% for support workers. The difference between these two position types, which are determined primarily by the nature of their work, suggests that tasks requiring deep thought and reflection may benefit from fewer interruptions~\cite{Golden2019-ff, Bernstein2018-is} when completed at home rather than in the office. However, at the higher end of the curves, there is no significant difference in turnover risk between fully remote workers in the two groups.

\subsubsection{Leader turnover risk is lower at low remote frequency but exceeds that of individual contributors at high frequency}
Bloom et al.~\cite{Bloom2024-yi} show that a hybrid arrangement is highly valued by employees, except for managers. At a remote frequency of two days per week, non-managers exhibit a significant reduction in turnover risk, whereas managers show an insignificant increase. These findings motivate us to investigate the different effects of varying remote frequency on the two groups of employees. As shown in Fig.~\ref{fig:april_dropout_leadership}, leaders are far less likely to turn over than individual workers when working onsite, likely because they are more connected to other people and teams in the company, perceive a stronger fit with their roles, and would sacrifice more if they left their jobs~\cite{mitchell2001people, saeed2022impact}. While both groups see reduced turnover risk at low to moderate remote frequencies, it is interesting to note that before the turning point at 1.3 days per week, the curve for leaders is almost parallel to that of individual contributors, indicating similar effects of remote work. The sharper increase in turnover risk beyond three days of remote work per week, however, suggests that leaders are more sensitive to higher remote work frequencies. 
The difference between the two roles can be explained by the additional challenges leaders may face in remote settings. When transitioning from onsite to remote work, leaders must adapt traditional leadership practices to virtual formats. For example, when shifting from face-to-face interactions to technology-mediated communication, enhanced communication skills~\cite{Ye2025-go} are needed to ensure clarity and avoid misunderstandings. Leaders are also responsible for maintaining team cohesion, such as keeping remote members engaged and ensuring that their direct reports feel included in the team. 
Meanwhile, individual contributors may be more reluctant to ask for feedback when remote~\cite{Emanuel2023-xz}. Additionally, when leaders work remotely, the in-person attendance of individual contributors tends to be lower~\cite{Charpignon2023-db}. 
These challenges, which may have a negligible impact when remote work frequency is low, can become more pronounced at higher remote frequencies, leading leaders to perceive fewer benefits and greater limitations~\cite{Kowalski2022-zm}.

\section{Effects of Remote Work Frequency on Promotion}

\begin{figure*}[t!]
\centering
    \begin{subfigure}[b]{0.3\textwidth}
         \centering
         \includegraphics[width=\textwidth]{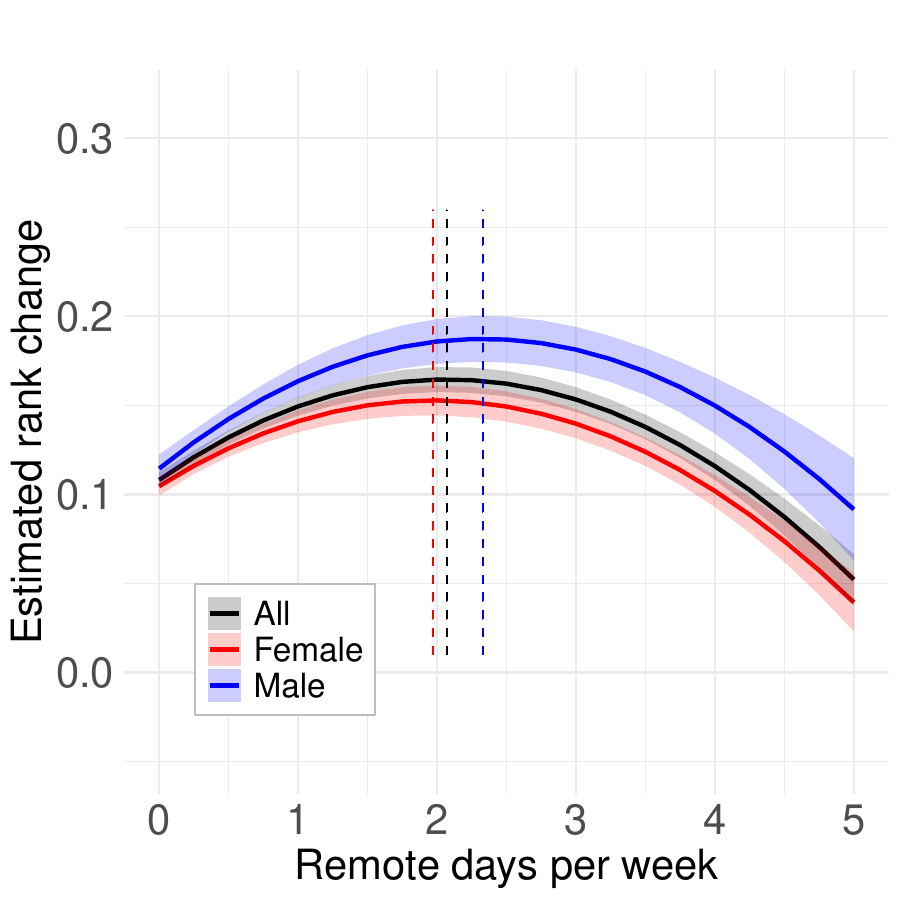}
         \caption{Gender}
         \label{fig:april_promotion_gender}
     \end{subfigure}
    \begin{subfigure}[b]{0.3\textwidth}
         \centering
         \includegraphics[width=\textwidth]{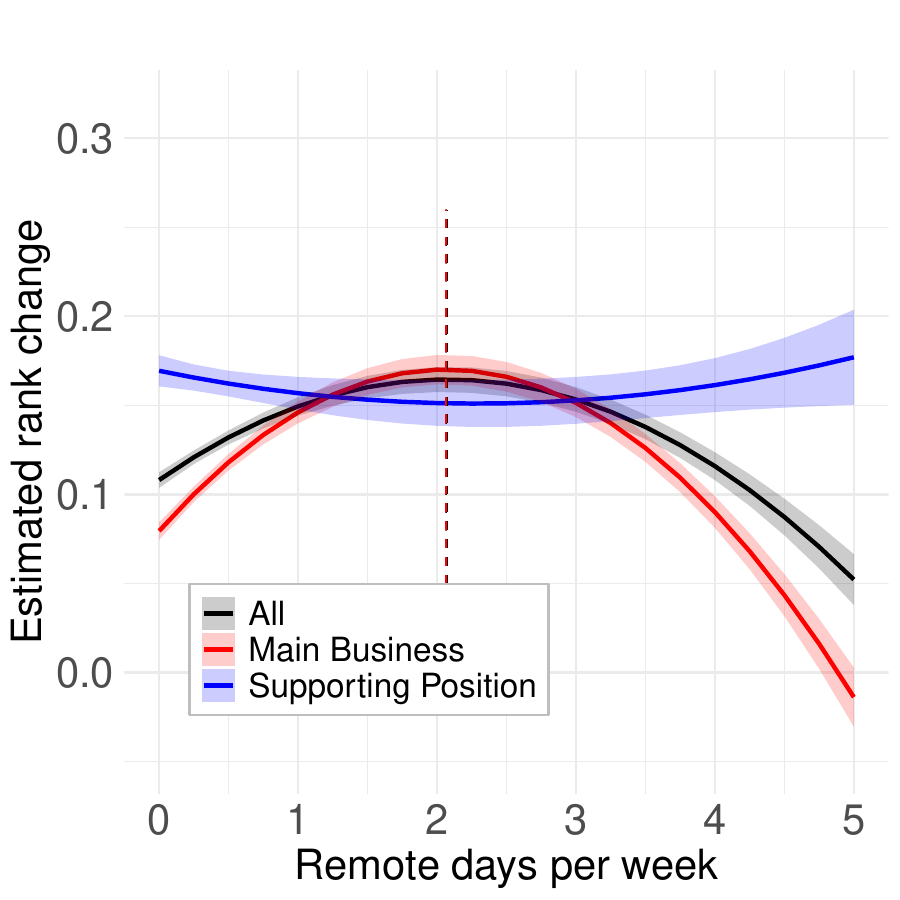}
         \caption{Position type}
         \label{fig:april_promotion_position}
     \end{subfigure}
     \begin{subfigure}[b]{0.3\textwidth}
         \centering
         \includegraphics[width=\textwidth]{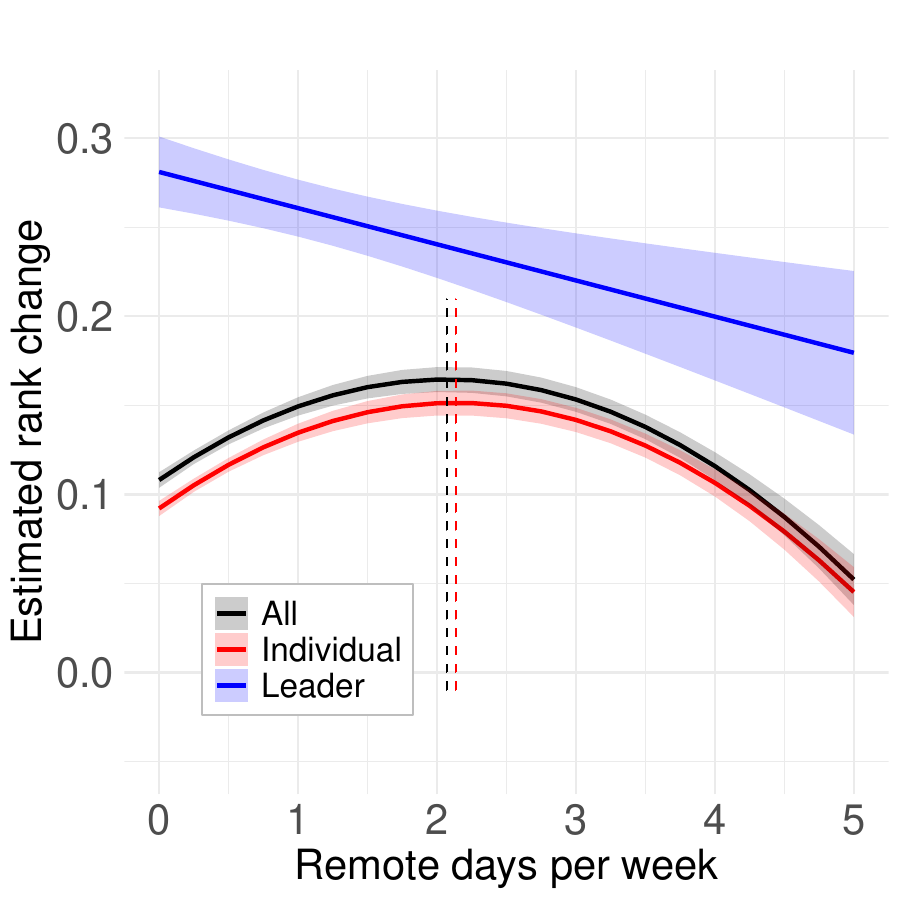}
         \caption{Leadership}
         \label{fig:april_promotion_leadership}
     \end{subfigure}
\caption{\textbf{Dose-response curves with 95\% confidence intervals for the weighted regressions of rank change on remote frequency.} As remote frequency increases, rank change initially rises, peaking at 2.1 remote days per week, and then begins to decline. Regressions on subgroups defined by gender, position type, and leadership status reveal heterogeneous effects. \textbf{(a)}: The optimal remote frequency for maximizing promotion is 2.3 days per week for men and 2.0 days per week for women. \textbf{(b)}: For main business workers, the optimal frequency is 2.1 days per week, while the effect for support workers is limited. \textbf{(c)}: For individual contributors, the optimal frequency is 2.3 days per week, while leaders achieve the highest promotion likelihood when working fully on-site. }\label{fig:april_promotion}
\end{figure*}

In this work, we use monthly changes in position rank (hereafter referred to as \textit{rank change}) to measure promotion. A positive rank change indicates a promotion, with larger changes reflecting more significant advancements, while a negative rank change indicates a demotion. For all employees with position rank information in May, the overall promotion rate and demotion rate are 10.1\% and 0.1\%, respectively.
  
To investigate whether remote frequency in April influences position rank change in May, we consider the same set of confounders and apply the same approach as in the previous section. Specifically, we fit a weighted OLS regression of the rank change on remote frequency, with a quadratic term included. Given all significant coefficients ($p < 0.001$), we visualize the relationship between rank change and remote frequency using a dose-response curve for the fitted regression. Table S2 in Appendix reports the regression results.

\subsection{Promotion likelihood increases with remote frequency up to 2.1 days per week, then declines}
The inverted U-shaped relationship between estimated rank change and remote frequency, as illustrated by
Fig.~\ref{fig:april_promotion}, indicates that remote work can actually enhance promotion likelihood within a certain range of frequencies. The rank change for fully onsite workers is 0.108 but can reach 0.165 for those who work remotely for 2.1 days per week, showing a 52.3\% increase. This increase may be due to increased flexibility for partially remote employees, which can contribute to higher productivity~\cite{Bloom2024-yi}, as well as more available time for work resulting from the elimination of commute time~\cite{bailey2002review, guimaraes1999empirically}. However, a higher frequency over 4.1 days a week shows negative impacts on promotion, and the rank change for full remote workers is 0.052, 51.7\% lower than fully onsite workers. This result indicates that workers may face career penalties when engaging in frequent remote work. Due to their lack of physical presence in the workplace, it becomes more difficult for their dedication and contributions to be visible and properly acknowledged in performance evaluations. Additionally, they may experience a significant flexibility stigma, which can negatively affect perceptions of their competence and commitment~\cite{munsch2014pluralistic}.

\subsection{Effects of remote work frequency on promotion are moderated by gender, position, and leadership}
Similarly, we fit regressions across groups defined by gender, position type, and leadership status to investigate how these characteristics may moderate the effects of remote work frequency on promotion. The inverted U-shaped pattern holds across genders, but the extent of benefits gained from remote work differs. Main business workers exhibit a clear inverted U-shaped relationship between remote work frequency and promotion, while supporting employees do not. For supporting employees, promotion follows a U-shaped pattern, although the change in rank advancement appears to be minor across different levels of remote work frequency. Leaders, unlike individual contributors, exhibit a negative linear relationship between remote frequency and rank change, suggesting that higher remote frequencies negatively impact their promotion prospects.

\subsubsection{Males gain more promotion benefit from remote work} Males and females exhibit similar changes in rank when working onsite, with males having slightly higher values in rank change. Interestingly, as shown in Fig.~\ref{fig:april_promotion_gender}, the gender difference becomes non-trivial as remote work frequency increases: males achieve their highest rank change of 0.187 at a remote frequency of 2.3 days per week, while females reach their peak rank change of 0.152 at 2.0 days. In other words, to maximize the benefits of remote work for positive rank change, females should work remotely for 2.0 days per week—about 3 hours (assuming an 8-hour workday) fewer than males. More notably, males can achieve a maximum increase of 0.073 (63.7\%) in rank change, compared to 0.048 (46.1\%) for females. Although both groups experience a decline in promotion likelihood with higher remote frequencies, males consistently exhibit higher rank changes than females. This finding suggests an exacerbated disparity in promotion likelihood between males and females in hybrid and fully remote settings, which may worsen the under-representation of women in leadership positions~\cite{Lawson2022-mx}.

\subsubsection{Effects of remote frequency on promotion are clear in main business positions but limited in support positions}  
As shown in Fig.~\ref{fig:april_promotion_position}, main business workers exhibit an inverted U-shaped relationship between rank change and remote frequency, with high frequencies (e.g., fully remote) potentially leading to demotion. Their rank changes can be doubled at 2.1 days of remote work per week compared to fully onsite work (114.1\% increase). For workers in supporting positions, although the quadratic relationship still exists, 
the wide opening of the curve and the broad confidence interval suggest that the effect of remote frequency is minimal. One possible reason is the lower competition for promotion in this group, as the overall promotion rate in supporting positions is 15.5\%, significantly higher than that in main business positions (7.4\%). 

\subsubsection{Leaders' promotion decreases with higher remote frequency} For leaders, rank change and remote frequency exhibit a significant linear relationship ($p<0.001$) rather than a quadratic one. This aligns with the insights from the turnover analysis, as leaders may face greater challenges at high remote frequencies, potentially leading to lower work evaluation ratings.

\section{Time Allocation Mediates Effects of Remote Frequency}

To understand the underlying reasons for the effects of remote frequency, we conduct a mediation analysis to examine whether and how workers’ time allocation plays a role, as it can be directly influenced by the intensity of remote work and is related to working status, (mental) health, and ultimately, career success. 

We measure time allocation along three dimensions: working hours, working time flexibility, and after-hours work. The working hours dimension is measured by the average number of working hours per working day. For working time flexibility, we  
measure \textit{daily schedule flexibility} by quantifying the Shannon entropy~\cite{shannon2001mathematical} of a worker’s daily hours. Specifically, we group working hours into bins and compute the entropy based on the probability distribution across these bins. A higher value of entropy indicates greater flexibility in daily working hours. For example, an individual may be equally likely to work 2–4 hours or 8–10 hours on a given day. In contrast, a lower value suggests relatively fixed daily working hours. To account for scheduling patterns across the seven days of the week, we assign working activity to each weekday (Monday through Sunday) and calculate the entropy to measure \textit{weekly schedule flexibility}. For after-hours work, we quantify \textit{late work} with the proportion of days with late working hours, and \textit{weekend work} with the proportion of weekend days among all working days with recorded activity.

To estimate the indirect effect of remote frequency through mediators introduced above, it is essential to first understand the relationship between the mediator and remote frequency, and the relationship between the outcome and the mediator. The data show quadratic relationships in both cases for all mediators, indicating that the indirect effect is not constant. Therefore, we use the \textit{instantaneous indirect effect}~\cite{hayes2010quantifying}—the rate at which changes in remote frequency indirectly affect outcome through changes in the mediator—for the mediation analysis. See~\nameref{sec:methods} for details. For clarity, we focus on remote frequencies in increments of 0.5 days in the following analysis.

\begin{figure}
    \centering
    \includegraphics[width=0.45\linewidth]{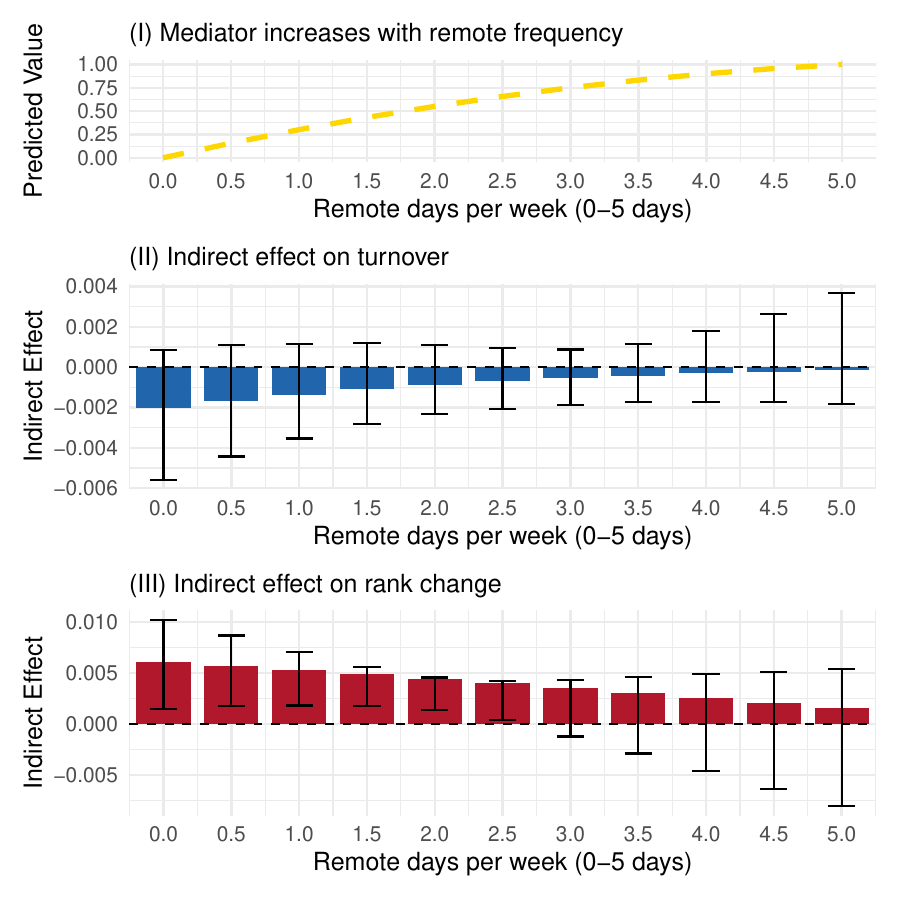}
    \caption{\textbf{Instantaneous indirect effects (with 95\% confidence interval) of remote frequency through working hours.} (I) Working hours (scaled and vertically offset for clarity) increase with remote frequency. (II) Working hours do not significantly mediate the effect of remote frequency on turnover. (III) Increasing working hours strengthen the positive effect of remote work on promotion when remote frequency is 2.5 days or fewer per week. At higher remote frequencies, the indirect effect becomes non-significant, even though working hours continue to increase. 
    }
    \label{fig:mediate_v3_hours}
\end{figure}

\subsection{Mediation through working hours}
As shown in Fig.~\ref{fig:mediate_v3_hours}, daily working hours increase with the number of remote days per week, but this variable shows no significant mediation effect on turnover risk. This is noteworthy, as working hours are typically regarded as an indicator of workload and, in that sense, should be related to job satisfaction~\cite{Schall2019-gn} and turnover risk. One possible reason is that, when employees work remotely, their longer daily working hours may include non-work activities, and their actual effort devoted to work may not change significantly. Regarding rank change, the increase in working hours shows a positive mediation effect for remote frequencies ranging from 0 to 2.5 days per week. As an indicator of motivation~\cite{Whitely1991-kt}, longer working hours may reflect greater dedication, contributing to more favorable evaluations of job performance. However, this effect disappears at higher levels of remote frequency, possibly due to negative perceptions or penalties directed at employees who use flexible work arrangements, even when their performance is unchanged or improved.~\cite{Golden2020-ju}.

\begin{figure}
    \centering
    \begin{subfigure}[b]{0.45\textwidth}
        \centering
         \includegraphics[width=\textwidth]{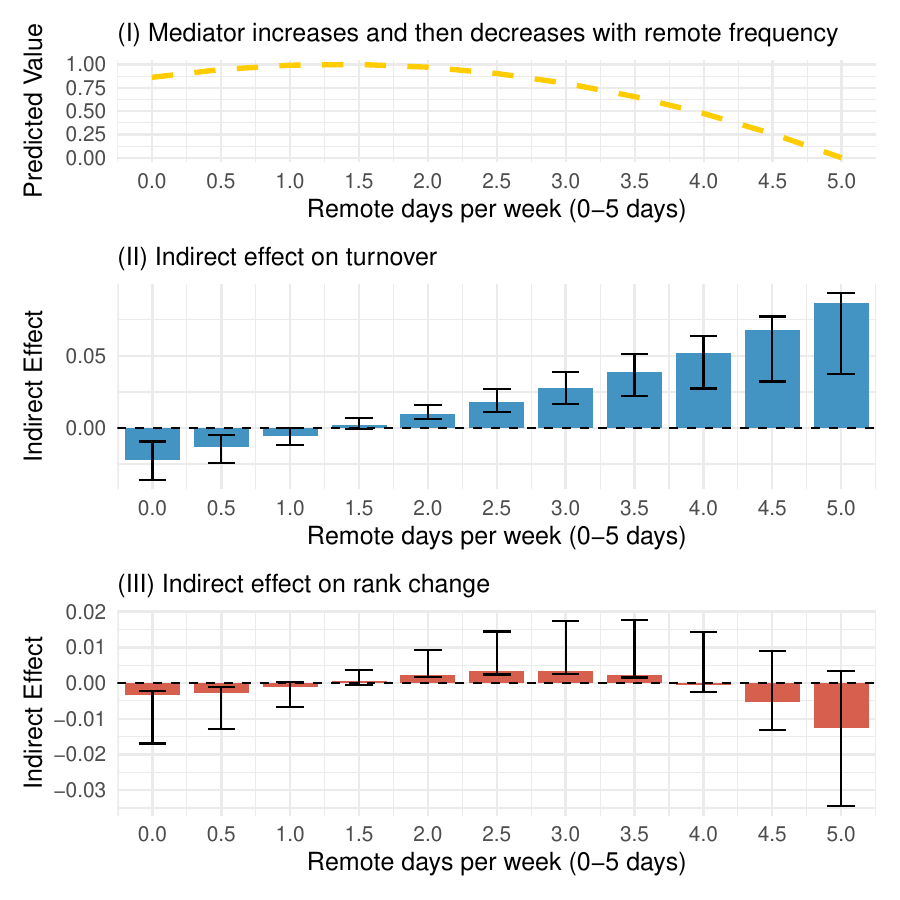}
         \caption{Weekly schedule flexibility}
         \label{fig:mediate_v3_flexibility_weekly}
    \end{subfigure}
    \begin{subfigure}[b]{0.45\textwidth}
        \centering
         \includegraphics[width=\textwidth]{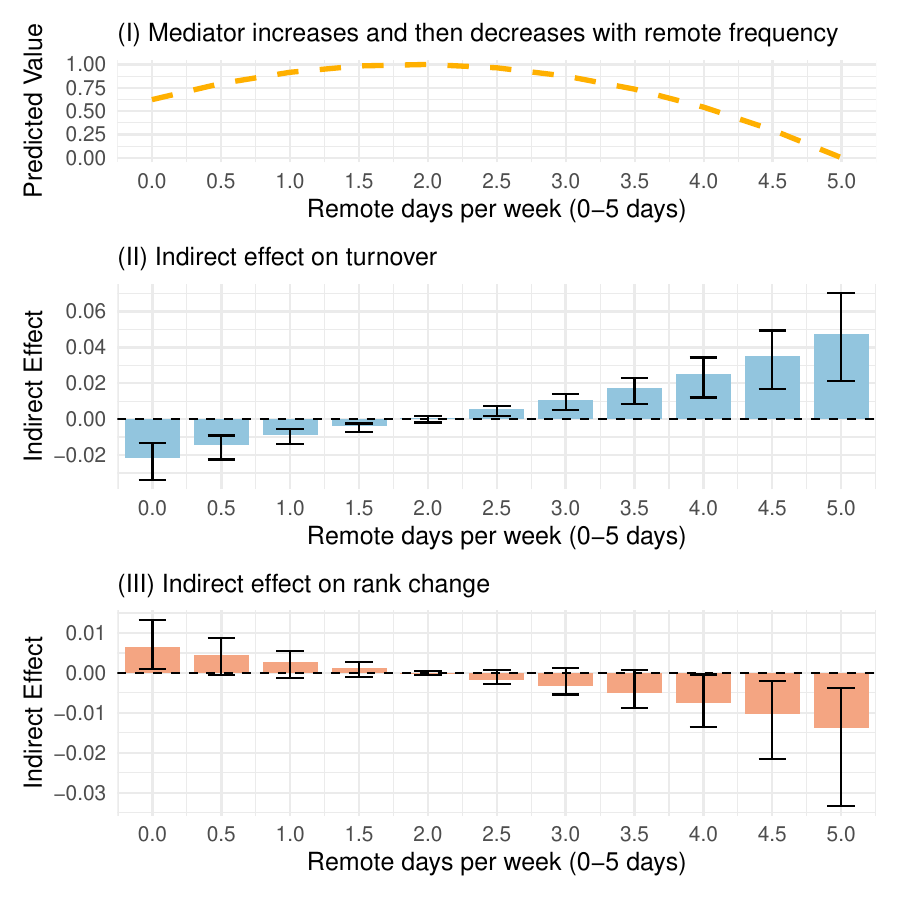}
         \caption{Daily schedule flexibility}
         \label{fig:mediate_v3_flexibility_daily}
    \end{subfigure}
    \caption{\textbf{Instantaneous indirect effects (with 95\% confidence interval) of remote frequency through flexibility.} \textbf{(a)}: (I) Weekly schedule flexibility increases with remote frequency up to 1.5 days per week, then decreases, reaching its lowest point under fully remote conditions. (II) Flexibility strengthens the effect of remote frequency on turnover: when remote frequency is below 1 day per week, increased flexibility helps reduce turnover risk; when remote frequency exceeds 1.5 days per week, decreasing flexibility contributes to higher turnover risk. (III) At low remote frequencies ($<$ 1.5 days/week), increased flexibility buffers the positive effect of remote work on promotion. At moderate frequencies (2–3.5 days/week), declining flexibility mediates a positive effect. At high frequencies (4–5 days/week), flexibility drops below onsite levels, and the mediation effect becomes non-significant. \textbf{(b)} (I) Daily schedule flexibility (measured by the entropy of working hours) increases with remote frequency up to 2 days per week, then decreases. (II) Flexibility strengthens the effect of remote frequency on turnover: increased flexibility reduces turnover risk at remote frequencies up to 1.5 days per week, while at frequencies above 2.5 days per week, decreasing flexibility contributes to reduced turnover. (III) Flexibility also strengthens the effect of remote frequency on promotion at both low and high ends. When remote frequency is below 0.5 days per week, increased flexibility positively mediates the effect of remote work; at high remote frequencies ($\geq$ 4 days/week), flexibility is even lower than fully onsite conditions and negatively mediates the effect of remote work.
    }
    \label{fig:mediate_v3_flexibility}
\end{figure}

\subsection{Mediation through flexibility}
The weekly schedule flexibility (Fig.~\ref{fig:mediate_v3_flexibility_weekly}) increases when remote frequency is low (less than 1.5 days per week), contributing to lower turnover risk. This suggests that, compared with the level of weekly schedule flexibility in fully onsite work, having greater flexibility to work on any day of the week is beneficial for retaining workers. However, when remote frequency exceeds 1.5 days per week, the weekly schedule flexibility begins to decrease and eventually falls below the level observed in fully onsite work. This indicates that workers tend to work on the same specific weekday(s), in a pattern even more condensed than fully onsite workers, which contributes to higher turnover risk. The mediation effect of this variable on rank change is quite different. When remote frequency is low (less than 1.5 days per week), the slightly increasing flexibility in weekly schedule negatively mediates its effect on rank change, though only marginally. When remote frequency falls within the range of 2 to 3.5 days per week, the decreasing flexibility shows a positive mediation effect. These findings suggest that, from the perspective of promotion, a more spread-out weekly schedule may not be favored—possibly due to increased communication costs~\cite{Brucks2022-tx}. At even higher levels of remote frequency, the mediation effects disappear, suggesting that fully remote workers may have already developed stable communication patterns with their teams.

The flexibility in daily schedule (Fig.~\ref{fig:mediate_v3_flexibility_daily}) has a positive relationship with remote frequency at low to moderate levels. For remote frequencies up to 1.5 days per week, an increase in flexibility negatively affects dropout risk, suggesting that greater flexibility in working hours provides benefits for employees. 
However, the decrease in flexibility at higher levels of remote frequency (more than 2.5 days per week) may contribute to higher turnover risk. Regarding rank change, daily schedule flexibility mediates the effects of remote frequency only among very infrequent remote workers (less than 0.5 days per week) and very frequent remote workers (at least 4 days per week). 
This result implies that performance evaluation is primarily sensitive to these two groups of employees. For those who seldom take advantage of remote work opportunities, increased flexibility in working hours may enhance their motivation and productivity. In contrast, for those who work remotely most of the time, a highly condensed daily schedule may become a negative indicator in job performance, possibly due to reduced flexibility in collaboration. 

Flexibility has long been considered one of the greatest benefits of remote work; however, our findings suggest that this relationship warrants re-examination. Each of the two work time flexibility-related mediators (i.e., weekly schedule flexibility and daily schedule flexibility) exhibits an inverted U-shaped relationship with remote frequency. As employees transition from fully onsite to a moderate level of remote work (around 1.5 to 2 days per week), their flexibility increases. However, when remote work becomes more intensive (e.g., more than three days per week), the values of these mediators begin to decline, eventually dropping below the levels observed for fully onsite workers, indicating reduced flexibility. Such observations indicate that workers tend to develop routines in higher remote frequency settings. In a fully remote setting, employees are expected to have complete flexibility in managing their time between work and personal life. However, this can actually lead to reduced flexibility in working hours, as non-work activities may take priority, narrowing the available days and hours for focused work. This dynamic may negatively affect remote workers’ perceived autonomy~\cite{deci2004handbook, jung2011understanding} with respect to scheduling, or the extent to which they feel they can control when they work.
Our findings help explain why a moderate level of remote work is most beneficial for career success: it offers the greatest flexibility for structuring work time while maintaining clear boundaries between work and personal life. 

\begin{figure}
    \centering
    \begin{subfigure}[b]{0.45\textwidth}
        \centering
         \includegraphics[width=\textwidth]{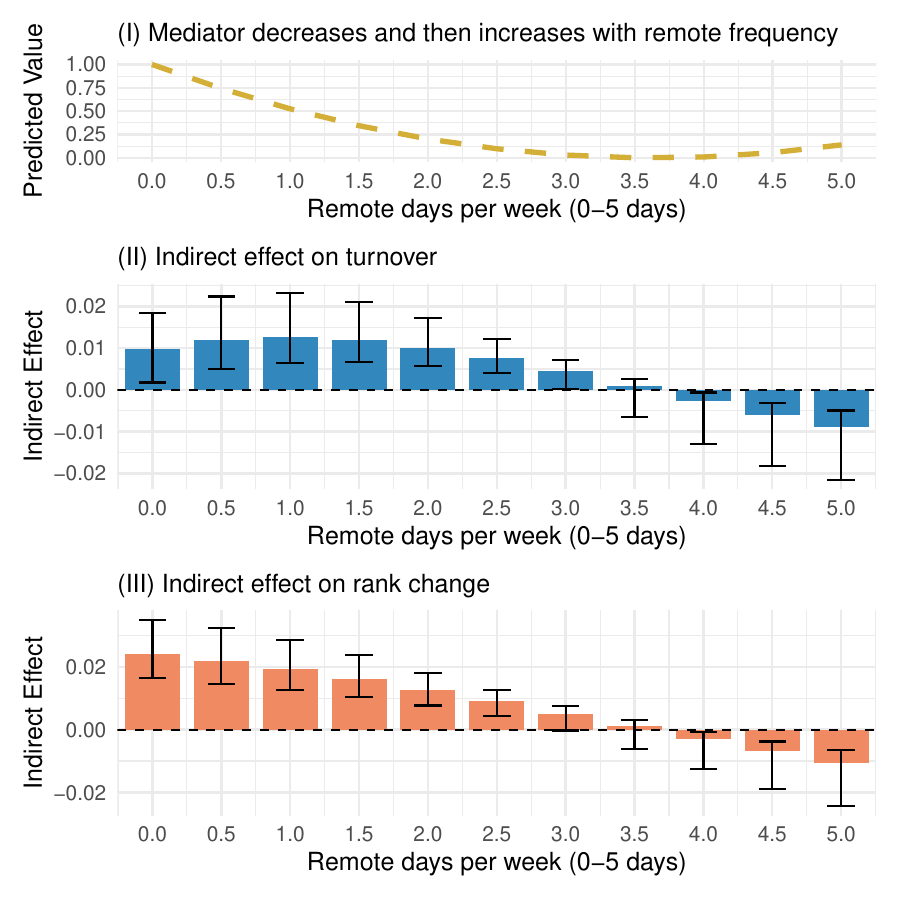}
         \caption{Weekend work}
    \end{subfigure}
    \begin{subfigure}[b]{0.45\textwidth}
        \centering
         \includegraphics[width=\textwidth]{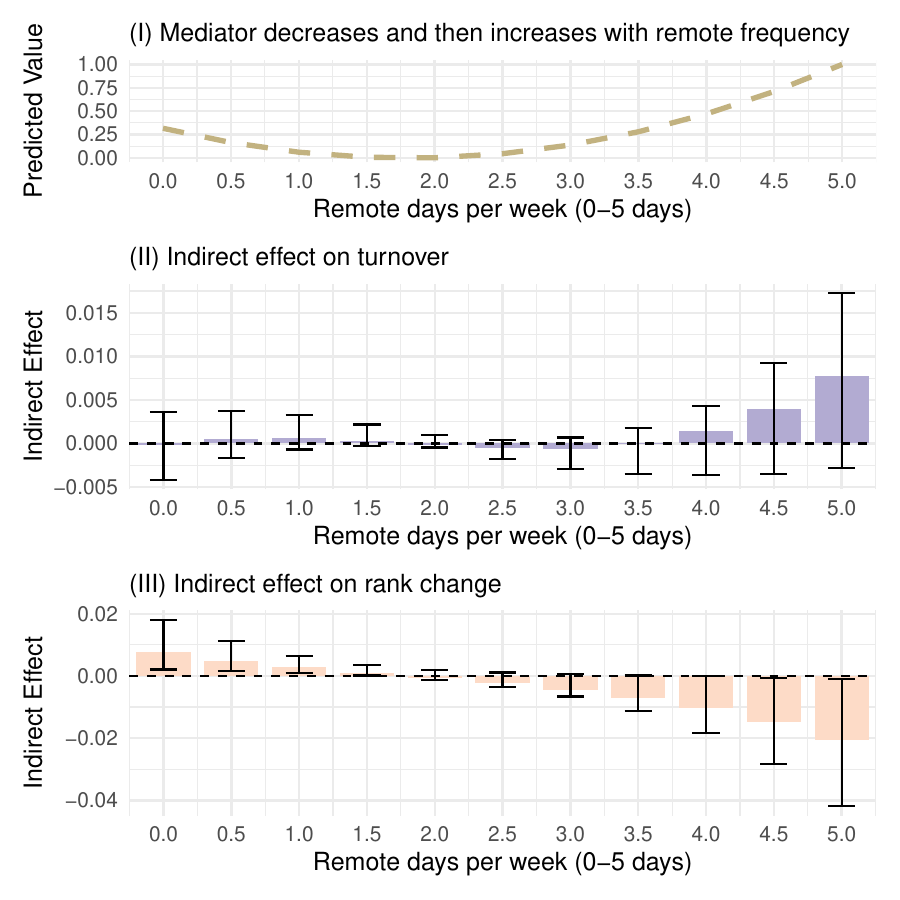}
         \caption{Late work}
    \end{subfigure}
    \caption{\textbf{Instantaneous indirect effects (with 95\% confidence interval) of remote frequency through after-hour work.} \textbf{(a)}: (I) Weekend work (measured by the proportion of working days falling on weekends) decreases with remote frequency until 3.5 days/week, then slightly increases. (II) Weekend work buffers the effect of remote work on turnover: for remote frequencies up to 3 days/week, reduced weekend work mediates a positive effect; at higher frequencies (4–5 days/week), a slight increase in weekend work helps reduce turnover risk. (III) Weekend work strengthens the effect on promotion: reduced weekend work ($\leq$ 3 days/week remote) contributes to a higher chance of promotion, while increased weekend work at higher remote frequencies shows a negative mediation effect. \textbf{(b)}: (I) Late work (measured by the proportion of working days with evening activity) decreases with remote frequency until 2 days/week, then increases. (II) Late work does not significantly mediate the effect of remote work on turnover. (III) Late work strengthens the effect on promotion: decreasing late work at lower remote frequencies ($<$ 1.5 days/week) positively mediates the promotion effect, while increasing late work at high remote frequencies ($\geq$ 4 days/week) results in a negative mediation effect. 
    }
    \label{fig:mediate_v3_after}
\end{figure}

\subsection{Mediation through after-hours work} 

As remote frequency increases from zero to three days per week, the weekend work  decreases, and the indirect effect of remote frequency on turnover through this mediator is positive. Although the weekend work lowers, having to work on weekends over a certain amount can still be a negative factor to retain workers. 
The proportion slightly increases after reaching its lowest point around 3.5 days per week, contributing to reduced turnover at higher remote frequencies of 4–5 days per week.
More frequent weekend work at high levels of remote frequency could signal strong motivation, or it may reflect mental compensation, as implied by exchange theory~\cite{blau2017exchange, Golden2019-ff}, intensive remote workers feel obligated to exert extra effort and diligence to justify the benefits of remote work. 
On the other hand, working on weekends appears to improve promotion opportunities at lower remote frequencies (up to 3 days per week), likely due to the greater presence and perceived dedication, which are positively associated with promotion evaluations.  
At higher remote frequencies (at least 4 days per week), the mediation effect becomes negative despite an increase in weekend work, suggesting that employees with high remote frequencies may need to make their efforts more visible to be considered for promotion. 

Late work shows no significant mediation effect on turnover, possibly for the same reason as working hours. However, this variable does show mediation effects on rank change, similar to the proportion of weekend work, with positive effects for remote frequencies below 1.5 days per week and negative effects above 4 days per week.

These two mediators represent how remote workers are motivated to do supplemental work~\cite{Golden2020-ju} in different ways. In general, as remote work frequency increases, workers tend to work less on weekends. Although the proportion of weekend work rises slightly at the highest levels of remote frequency, it remains much lower than that of fully onsite workers. The proportion of late work, while also exhibiting a U-shaped relationship, conveys a clearer message. Late work increases significantly at higher levels of remote frequency, surpassing the levels observed among fully onsite workers, yet it still negatively mediates the effect of remote frequency on promotion. This further illustrates the bias and flexibility stigma~\cite{Golden2020-ju} that remote workers may face, suggesting that frequent remote workers may need to make additional efforts to signal their dedication to managers~\cite{Afota2022-ma}.

\section{Discussion}

Given the different nature of turnover and promotion, is there an optimal frequency of remote work that benefits both of them? Compared to fully onsite workers and those with intensive remote arrangements, our results show that working remotely for up to three days per week is associated with lower turnover risk, with the lowest risk observed at 1.6 days. Similarly, working remotely for up to four days per week can benefit promotion, with the largest positive rank change occurring at 2.1 days. Although the two outcomes may differ in priority in practice, our analysis suggests a beneficial remote frequency of up to three days per week for both outcomes, and an optimal remote frequency of two days per week (within the range of 1.6 to 2.1 days). Our analysis also reveals potential mechanisms underlying this optimal remote schedule in terms of work time allocation. Compared to fully on-site and intensive remote workers, two days of remote work are associated with a moderate level of working hours, which may strike a balance between high motivation~\cite{Whitely1991-kt} and the risk of invisibility when off-site~\cite{Golden2020-ju}. In addition, two days of remote work could link to greater flexibility in working time, which supports higher perceived autonomy~\cite{deci2004handbook, jung2011understanding} and a healthier work–life balance with clearer boundaries~\cite{Hayashi2024-lp}.

We also find these optimal effects of remote frequency can vary across groups defined by gender, position type, and leadership status, and an optimal frequency does not always exist. For men, the optimal remote frequency associated with the lowest turnover risk is 1.4 days per week, while the frequency associated with the highest promotion likelihood is 2.3 days. For women, the corresponding values are 1.8 and 2.0 days, respectively. 
These results suggest that two days per week can still be considered an optimal frequency across genders. 
However, these suggestions do not hold for all position types. Two days per week appears optimal for main business workers, who show the lowest turnover risk at 1.8 days and the highest promotion gains at 2.1 days. By contrast, support workers experience the lowest turnover risk at 1.1 days per week but do not benefit from remote work in terms of promotion. For leaders, the lowest turnover risk occurs at 1.3 days per week, and remote work has a negative effect on their promotion prospects. Our results suggest the importance of considering position characteristics (such as position type and leadership status) when designing effective remote work policies.

In addition to the optimal ranges of remote frequency for men and women, we observe notable gender differences in the effects of remote work. 
First, men’s turnover risk surpasses that of women at around two days per week, after which it rises sharply, reaching 55.5\% for fully remote work, compared to 33.2\% for women (see Fig.~\ref{fig:april_dropout_gender}). This substantial difference suggests that men and women may face distinct challenges when working remotely at high frequencies, which warrants further research. One possible explanation is that women experience greater time stress than men~\cite{craig2017feeling, bianchi2003gender, Whillans2021-uq}, which may lead them to remain more focused on work when remote, whereas men may devote more time to personal life and leisure. Second, in terms of promotion, men gain more benefits than women (see Fig.~\ref{fig:april_promotion_gender}), achieving a maximum rank change of 63.7\% compared to 46.1\% for women. At the same remote frequency, men show a higher rank increase than women, and the gap is larger than in fully on-site work. 
Does this gap exist because that women are less likely to enter competitions~\cite{He2021-ki}? Does remote work increase the likelihood of social exclusion for women~\cite{Cyr2021-ur}? The diversity of the leadership team is important for a company's innovativeness and sustainability~\cite{Hakovirta2023-xy}, and hiring women into senior leadership positions can help reduce gender stereotypes within the organization's culture~\cite{Lawson2022-mx, Ye2025-go}. Our findings highlight the need for future research to address such questions in order to reduce gender imbalances in leadership positions~\cite{He2021-ki} and promote gender equality~\cite{England2020-jy} in the workplace.

Our analysis conveys a clear message: a certain level of hybrid work (around two days per week) has advantages over both fully on-site work and frequent remote work (four to five days per week), optimizing turnover risk and promotion likelihood. In addition to the possible mechanisms we have discussed (i.e., working hours, flexibility, and after-hour work), an appropriate frequency of remote work can also enhance productivity~\cite{Monteiro2019-rn}, perceived autonomy~\cite{Gajendran2024-be}, job satisfaction~\cite{Schall2019-gn}, the generation of creative ideas~\cite{Brucks2022-tx} and breakthrough discoveries~\cite{lin2023remote}, as well as collective performance in solving complex problems~\cite{Bernstein2018-is}. Requiring full onsite attendance would cause companies and employees to lose these benefits. Allowing a high frequency of remote work, or even fully remote arrangements, may lead to higher turnover risks and lower promotion opportunities, as higher frequencies of remote work are associated with longer working hours, reduced flexibility in time allocation, and more late work, which in turn can contribute to other adverse work-related outcomes and mental health issues~\cite{du-Plooy-and-Gert-Roodt2010-zs}.

This work contributes to the literature in the following ways. First, we provide empirical evidence on how remote work frequency influences work-related outcomes using large-scale, real-world behavioral data, complementing prior research based on surveys and interviews. Second, our results on the effects of remote frequency offer implications for experimental design, where identifying an optimal treatment level is rarely feasible because setting multiple treatment levels is costly. Third, the analyses by gender, position type, and leadership status, along with the examination of underlying mechanisms through time allocation, suggest promising directions for future research.

This work provides practical implications for various stakeholders. For policymakers, our findings suggest encouraging hybrid work arrangements. Mandatory return-to-office requirements may harm work-related outcomes (for example, increased turnover rates) by imposing more rigid schedules and longer after-hours work, which can lead to worse work-life balance and lower motivation. For companies and managers, lower turnover rates and higher job satisfaction can be achieved by identifying the optimal range of remote-work frequency. Our results indicate that two remote days per week is optimal, aligning with the most common hybrid pattern~\cite{aksoy2023working, barrero2023evolution}. A customized hybrid arrangement could also be helpful when position characteristics are taken into account. For example, leaders may benefit from reducing their remote-work frequency, which could lower turnover risk and improve promotion likelihood. Additionally, our findings indicate that measures should be taken to ensure effective communication and strong engagement among remote workers. Companies should also emphasize gender equity in promotion policies, as women are more disadvantaged when working remotely, and take measures to increase women’s visibility~\cite{He2021-ki} and participation in competitive opportunities~\cite{He2021-ki}. For workers, our mediation analysis illustrates how working patterns and work–life balance~\cite{Sullivan2012-zj, Albertsen2008-lb} are associated with remote-work frequency. When the frequency is fixed, workers may adopt personalized strategies to enhance the benefits of remote work. For instance, intensive remote workers with low daily flexibility for work may establish clearer work-life boundaries and improve accessibility for collaboration.

In summary, this study demonstrates that the optimal hybrid arrangement involves two days of remote work per week. The design of our analysis is constrained by data limitations.  
First, the outcomes of career success are measured using short-term data (i.e., outcomes within a month) and may therefore neglect factors that affect long-term outcomes. This limitation stems from the dataset and the context of data collection (see~\nameref{sec:methods}). 
Although our results do not indicate long-term effects of remote work, the mediation analysis provides valuable findings on how time allocation varies with changes in remote frequency and how it helps explain the observed effects, thereby offering actionable insights for employees, organizations, and even governments.
Second, as our analysis is based on data from a single company in one country, generalization to other settings~\cite{Aksoy2025-pr, Bartik2020-tp} should be made with caution. 
Finally, although some confounders that may influence both remote work frequency and career success~\cite{Mariani2024-ez} cannot be measured due to data limitations, we accounted for as many confounders as possible and ensured that the covariates are well balanced, indicating a reasonable replication of a randomized experimental setting. 
Despite these inherent limitations, the dataset remains valuable because it provides fine-grained attendance records and position rank data for a large number of workers, enabling an in-depth analysis of the effects of remote work. More importantly, such diverse remote work patterns are rarely found within a single company, as organizations typically adopt unified remote work strategies. 
For future work, we plan to expand the time range of our data to investigate the effects of external shocks (such as COVID-19 and the Double Reduction Policy) on how remote work influences career outcomes. Once the effects of external shocks are well captured, we will examine the impact of remote work frequency over longer periods. In addition, we plan to incorporate more work-related activities (such as document editing and video conferencing) to study social behaviors~\cite{Napoli2023-nq} and communication networks~\cite{Yang2021-nn, Yang2022-ki} under different remote work settings.

\begin{small}
\section*{Materials and Methods} 

\label{sec:methods}

\subsection*{Data selection}\label{subsec:data}
We use a dataset collected from an education company that includes employee daily attendance activities and monthly position rankings in the year of 2021. The company ranks among the top 500 in China and had over 100,000 employees during the data collection period. We choose to investigate whether and how remote frequency in April 2021 influenced employees’ career success in the following month for three reasons. 
First, our raw dataset includes attendance records spanning the entire year of 2021, while position rank data is only available starting from April. 
Second, our dataset was collected during the COVID-19 pandemic. The employees in our dataset were based in Beijing, where a city-wide travel restriction was in place until March 2021, and a new control was implemented in July 2021. To minimize the impact of the pandemic on working patterns, we need to focus on remote work activities that occurred after March and before July. 
Third, the education company (along with the broader education industry) was significantly affected by the Double Reduction Policy\footnote{A Chinese education policy issued on July 24, 2021 aimed to reduce homework and after-school tutoring burdens for primary and secondary school students, and was directly related to the core business of the company we studied. See \url{https://en.wikipedia.org/wiki/Double_Reduction_Policy}.} starting in July 2021. Notably, our dataset shows a sharp increase in the dropout rate to 22.8\% in July, up from 12.3\% in May and 10.8\% in June, indicating that a substantial number of employees may have left the company since July, possibly in response to the policy. Finally, the proportion of employees who received a rank change in June is only 0.2\%, compared with 10.1\% in May. The limited number of promotions restricts the analysis of the effects of remote work on this outcome. All of these factors constrained the timing of our study, leaving only one viable option: using working patterns from April, when control measures were relatively relaxed, and career outcomes from May, before they were affected by the subsequent policy shock.  
In the analysis, we select employees with work records in both March and April, resulting in a final sample size of 41,579. The positions are categorized into main business (i.e., teaching and research, n = 27,519) and support (n = 14,060, including marketing, operations, sales, and technical support).

\subsection*{Remote work frequency measurement}\label{subsec:freq}
In literature, remote work frequency (or the intensity of remote work) is measured at different scales, including the average number of hours working remotely~\cite{Golden2005-xr} and the number of days working remotely per week measured on a six-point Likert scale reported by survey respondents, where 0 represents less than one day per week or none and six means five days per week~\cite{Hayashi2024-lp}. In this work, we calculate the remote work frequency using records of workers' attendance activities. Specifically, for a given month, remote frequency is measured as the number of days an employee works remotely divided by the total days they worked, normalized to a five-day workweek. This frequency is an approximately continuous variable ranging from 0 to 5. 

\subsection*{Modeling relationship between remote work frequency and career outcomes}
To move beyond simple correlations, we adopt the potential outcome framework~\cite{Rubin2005-pe} to estimate the effects of remote work frequency using observational data. In particular, since the ``treatment,'' the remote frequency, is an approximately continuous variable, we use the Generalized Propensity Score (GPS)~\cite{Hirano2005-hj} to quantify the probability that an individual with given confounders is exposed to a specific level of treatment.
The generalized propensity scores are estimated by regressing the continuous treatment (remote frequency) with a machine learning method called the gradient boosting machine~\cite{chen2016xgboost, zhu2015boosting}. Then, GPS is used to construct a pseudo-population, with the weights defined as $f(remote\_frequency)/GPS$~\cite{robins2000marginal}, where $f$ is the marginal density function. We apply a trimming strategy to minimize the influence of outlier weights. The covariate balance on the weighted pseudo-population is ensured through average absolute correlations (AC)~\cite{Wu2018-bg, Austin2019-oj}, which indicates good balance if the value is less than 0.1.  
We then fit an Ordinary Least Squares (OLS) model regressing the outcome (turnover and rank change) on the frequency of remote work including quadratic terms, incorporating the weights. A dose-response curve is used to visualize the relationship between the treatment variable (remote frequency) and the outcome variable with a 95\% confidence interval. 

\subsection*{Confounders}
The confounders consist of three components: working experience, working status, and structural roles, as these factors can potentially influence both the remote frequency of workers and their career success outcomes.
\begin{itemize}
    \item Work experience is measured by job tenure, i.e., the number of years an employee has worked in the organization. Job tenure and total time in one's occupation are positively associated with career success~\cite{cox1991career, Whitely1991-kt}. Meanwhile, job tenure can potentially affect one's frequency of remote work, as junior workers tend to work onsite more.
    \item Work status describes the workload and work activities of individuals and is assessed using data from the previous month to mitigate reverse causality~\cite{granger1969investigating}, as remote frequency and work status within the same month might interact. The measurements of working status include the number of working days and the average working hours per day in the previous month. Additionally, remote frequency in the previous month is included to control for other potential confounders of remote work frequencies, as well as the impact of prior remote work activities.
    \item Structural roles may influence both remote work decisions and expectations for job performance, which in turn can affect outcomes such as turnover and promotion. Structural roles are measured in two dimensions: (1) the official rank of the position and (2) leadership status, which is captured by the number of direct reports, the total reporting span, and the reporting depth below the worker. The reporting depth is defined as 0 if a worker has no direct reports.
\end{itemize}

\subsection*{Mediation Analysis}
To understand the underlying mechanisms of the effects, we perform mediation analysis considering time allocations that could be directly affected by remote work frequency while also potentially influential on career success outcomes. In particular, as nonlinear relations are identified between potential mediators and treatment, we adopt instantaneous indirect effect~\cite{hayes2010quantifying} to test and quantify the mediation. More specifically, let $X$ denote the treatment, $M$ denote the mediator, and $Y$ denote the outcome. The instantaneous indirect effect quantifies the effect of $X$ on $Y$ through $M$ at a specific value $X=x$. In our dataset, the relationship between treatment and mediator can be formed as $\hat{M}=i_{1}+a_{1}X+a_{2}X^{2}$ and the relationship between outcome and mediator can be formed as $\hat{Y}=i_{2}+f(X)+b_{1}M+b_{2}M^{2}$. According to Table 1 in~\cite{hayes2010quantifying}, the Instantaneous Indirect Effect ($\theta$) of $X$ on $Y$ through $M$ is $(a_{1}+2a_{2}X)(b_{1}+2b_{2}\hat{M})$. For each mediator, we use the whole dataset to generate the $\theta$ value 
for each $x$ and use bootstrapping~\cite{hayes2010quantifying} to generate 95\% confidence intervals, which are used to determine the significance of the mediation effects.
\end{small}

\bibliography{tal}

\appendix
\renewcommand{\thetable}{S\arabic{table}}
\section*{Appendix}
We provide the regression results in the appendix. Table~\ref{tab:turnover_reg_summary} presents the results of regressions of turnover risk on remote frequency for all workers, as well as for subsets categorized by gender, position type, and leadership status. All quadratic and linear terms are significant at $p<0.001$. Table~\ref{tab:rank_reg_summary} shows the results of regressions of rank change on the same datasets, with some workers excluded due to missing rank change data. All quadratic and linear terms are significant at $p<0.001$ for the total sample, male workers, female workers, main business workers, and individual workers. For supporting workers, the quadratic relationship is less significant ($p<0.05$ for both quadratic and linear terms). For leaders, the quadratic relationship is not significant; instead, rank change exhibits a significant linear relationship with remote frequency ($\beta_{\text{remote\_frequency}}=-0.0203$, $p<0.001$).

\begin{table}[htbp]
\centering
\caption{Weighted linear regression of turnover risks on remote frequency across seven datasets.}
\label{tab:turnover_reg_summary}
\begin{tabular}{lrrrrrrrr}
\toprule
\textbf{Dataset} & $\beta$(Intercept) & SE & $\beta$(treat$^2$) & SE & $\beta$(treat) & SE & $R^2$ & $N$ \\
\midrule
All & 0.1498*** & (0.0022) & 0.0289*** & (0.0009) & -0.0938*** & (0.0038) & 0.035 & 41,578 \\
Male & 0.1350*** & (0.0037) & 0.0374*** & (0.0015) & -0.1032*** & (0.0063) & 0.0816 & 14,085 \\
Female & 0.1575*** & (0.0027) & 0.0245*** & (0.0011) & -0.0879*** & (0.0047) & 0.0206 & 27,493  \\
Main business & 0.1689*** & (0.0027) & 0.0326*** & (0.0012) & -0.1155*** & (0.0049) & 0.0318 & 27,519 \\
Supporting & 0.1094*** & (0.0038) & 0.0209*** & (0.0014) & -0.0475*** & (0.0060) & 0.0481 & 14,059  \\
Individual & 0.1547*** & (0.0023) & 0.0282*** & (0.0009) & -0.0930*** & (0.0040) & 0.0313 & 37,702  \\
Leader & 0.0966*** & (0.0066) & 0.0322*** & (0.0024) & -0.0868*** & (0.0102) & 0.0906 & 3,876  \\
\bottomrule
\end{tabular}
\\[3pt]
\parbox{0.95\linewidth}{\small
\textit{Notes:} \textit{Treat} refers to remote work frequency. Standard errors in parentheses. ***$p$ $<$ 0.001, **$p$ $<$ 0.01, *$p$ $<$ 0.05.}
\end{table}

\begin{table}[htbp]
\centering
\caption{Weighted linear regression of rank change on remote frequency across seven datasets.}
\label{tab:rank_reg_summary}
\begin{tabular}{lrrrrrrrr}
\toprule
\textbf{Dataset} & $\beta$(Intercept) & SE & $\beta$(treat$^2$) & SE & $\beta$(treat) & SE & $R^2$ & $N$ \\
\midrule
All & 0.1081*** & (0.0022) & -0.0131*** & (0.0009) & 0.0545*** & (0.0039) & 0.0052 & 40,166 \\
Male & 0.1145*** & (0.0041) & -0.0134*** & (0.0017) & 0.0626*** & (0.0070) & 0.0060 & 13,583 \\
Female & 0.1046*** & (0.0027) & -0.0124*** & (0.0011) & 0.0488*** & (0.0046) & 0.0050 & 26,583  \\
Main business & 0.0795*** & (0.0025) & -0.0213*** & (0.0011) & 0.0879*** & (0.0045) & 0.0147 & 26,843 \\
Supporting & 0.1694*** & (0.0045) & 0.0035* & (0.0017) & -0.0161* & (0.0073) & 0.0004 & 13,323 \\
Individual & 0.0921*** & (0.0022) & -0.0130*** & (0.0009) & 0.0555*** & (0.0038) & 0.0060 & 36,331  \\
Leader & 0.2744*** & (0.0111) & -0.0059 & (0.0041) & 0.0039 & (0.0174) & 0.0040 & 3,835  \\
\bottomrule
\end{tabular}
\\[3pt]
\parbox{0.95\linewidth}{\small
\textit{Notes:} \textit{Treat} refers to remote work frequency. Standard errors in parentheses. ***$p$ $<$ 0.001, **$p$ $<$ 0.01, *$p$ $<$ 0.05.}
\end{table}

	
\end{document}